\documentclass[a4paper,11pt]{article}
\usepackage{jheppub}
\usepackage[T1]{fontenc}
\usepackage{amsfonts}
\usepackage{amsmath}
\usepackage{amssymb,epsf}
\usepackage{latexsym}
\usepackage{graphicx,epsfig}
\usepackage{amssymb}
\usepackage{subfigure}
\usepackage{epstopdf}
\graphicspath{{Images/}}

\title{Critical behaviour of Lifshitz dilaton black holes}

\author[a]{Zeinab Dayyani}
\author[a,b]{Ahmad Sheykhi}
\affiliation[a]{Physics Department and Biruni Observatory, College
of Sciences, Shiraz University, Shiraz 71454, Iran}
\affiliation[b]{Research Institute for Astrophysics and Astronomy
of Maragha (RIAAM), P.O. Box 55134-441, Maragha, Iran}
\emailAdd{asheykhi@shirazu.ac.ir}

\abstract{Till now, critical behaviour of Lifshitz black holes, in
an extended $P-v$ space, has not been studied, because it is
impossible to find an analytical equation of state, $P=P(v,T)$,
for an arbitrary Lifshitz exponent $z$. In this paper, we adopt  a
new approach toward thermodynamic phase space and successfully
explore the critical behaviour of $(n+1)$-dimensional Lifshitz
dilaton black holes. For this purpose, we write down the equation
of state as $Q^s=Q^s(T,\Psi)$ with $\Psi=\left({\partial
M}/{\partial Q^{s} }\right)_{S,P}$ is the conjugate of $Q^s$ and
construct Smarr relation based on this new phase space as $
M=M(S,Q^{s},P)$, where $s=2p/(2p-1)$ with $p$ is the power of the
power-law Maxwell Lagrangian. We justify such a choice
mathematically and show that with this new phase space, the system
admits the critical behaviour and resembles the Van der Waals
fluid system when the cosmological constant (pressure) is treated
as a fixed parameter, while the charge of the system varies. We
obtain Gibbs free energy of the system and find swallow tail shape
in Gibbs diagrams which represents the first order phase
transition. Finally, we calculate the critical exponents and show
that although thermodynamic quantities depend on the metric
parameters such as $z$ , $p$ and $n$, the critical exponents are
the same as Van der Walls fluid-gas system. This alternative
viewpoint toward phase space of lifshitz dilaton black hole can be
understood easily since one can imagine such a change for a given
single black hole i.e., acquiring charge which induces the phase
transition. Our results further support the viewpoint suggested in
\cite{Dehy}.}

\begin{document}
\maketitle

\flushbottom

\section{Introduction}
Historically, Maldacena was the first who suggested, two decades
ago, the correspondence between gravity in an Anti-de Sitter (AdS)
spacetime and the Conformal Field Theory (CFT) living on the
boundary of spacetime known as  AdS/CFT correspondence
\cite{Mald}. According to Maldacena's conjecture the effects of
the string theory in a $d$-dimensional $AdS_{n+1}\times S^{d-n-1}$
spacetime can be appeared in the form of a field theory on an
$n$-dimensional $r$-constant brane which is the boundary of
$AdS_{n+1}$ spacetime. This idea has attracted a lot of enthusiasm
and has been investigated from various point of view \cite{AdS}.
The metric of the AdS spacetime is given by
\begin{equation}
ds^{2}=-\frac{r^{2}}{l^{2}}dt^{2}+\frac{l^{2}}{r^{2}}dr^{2}+r^{2}\sum%
\limits_{i=1}^{n-1}dx_{i}^{2},
\end{equation}%
which is invariant under an isotropic conformal transformation as follows:
\begin{equation}
t\rightarrow \lambda t,\text{ \ \ \ \ }x_{i}\rightarrow \lambda
x_{i},\text{ \ \ \ \ }r\rightarrow \lambda ^{-1}r.
\end{equation}%
On the other hand, the application of AdS/CFT is restricted to systems respected isotropic
scale invariance and quantum critical systems show scaling symmetry as
\begin{equation}
t\rightarrow \lambda ^{z}t,\text{ \ \ \ \ }x_{i}\rightarrow \lambda x_{i},%
\text{ \ \ \ \ }r\rightarrow \lambda ^{-1}r,
\end{equation}%
where $z$ is a dynamical critical exponent and is restricted as
$z>1$. This parameter shows the degrees of anisotropy between
space and time.  The Lifshitz spacetime was first introduced in
\cite{Lif,Mann} as
\begin{equation}
ds^{2}=-\frac{r^{2z}}{l^{2z}}dt^{2}+{\frac{l^{2}dr^{2}}{r^{2}}}%
+r^{2}\sum\limits_{i=1}^{n-1}dx_{i}^{2}.
\end{equation}%
The Lifshitz spacetime is not a vacuum solution of Einstein
gravity and so needs matter source. Usually a massive gauge field
plays the role of this matter source but it is nearly impossible
to obtain an analytic solution for arbitrary $z$ in such models.
As shown in Ref. \cite{tarrio} considering a dilaton field,
instead of a massive gauge field, can lead to exact analytical
solutions in Lifshitz spacetime (see also \cite{Kord}). Another
motivation is that string theory in its low energy limit reduces
to Einstein gravity with a scalar dilaton field coupled to gravity
and other fields \cite{dilaton}.

On the other hand, the studies on the critical behavior of black
holes have got a lot of attentions in a wide range of gravity
theories. For example, critical behavior of charged AdS black
holes has been studied in \cite{MannRN} and the author completed
the analogy between Reissner-Nordstrom-AdS black holes with the
Van der Walls liquid-gas system, with the same critical exponents.
The key assumption is to enlarge the thermodynamic phase space to
include the cosmological constant as a thermodynamic pressure and
its conjugate quantity as a thermodynamic volume
\cite{Do1,Ka,Do2,Do3,Ce1,Ur}. When the gauge field is the
Born-Infeld nonlinear electrodynamics, one needs more extended
phase space to introduce a new thermodynamic quantity conjugate to
the Born-Infeld parameter which is necessary for consistency of
both the first law of thermodynamics and the corresponding Smarr
relation \cite{MannBI}. Treating the cosmological constant as a
thermodynamic pressure, thermodynamics and $P-v$ criticality of
black holes in an extended phase space in the presence of
power-Maxwell \cite{HV} and exponential nonlinear electrodynamics
\cite{Hendi1} have been explored. The studies were also
generalized to other gravity theories. In this regards, the phase
structure of asymptotically AdS black holes with higher curvature
corrections such as Gauss-Bonnet \cite{GB1,GB2} and Lovelock
gravity \cite{Lovelock} have also been investigated. The studies
were also extended to the rotating black holes, where phase
transition and critical behavior of Myers-Perry black holes have
been investigated \cite{Sherkat}. Other studies on the critical
behavior of black hole spacetimes in an extended phase space have
been carried out in \cite{Sherkat1,Rabin,Zou,John,Amin2}.

Critical behavior of the Einstein-Maxwell-dilaton black holes has
been studied in \cite{Kamrani}. When the gauge field is in the
form of Born-Infeld \cite{Dayyani1} and power-Maxwell
\cite{Dayyani2} field, critical behavior of $(n+1)$-dimensional
dilaton black holes in an extended phase space have been
investigated. Taking into account the dilaton field in the
presence of logarithmic and exponential forms of nonlinear
electrodynamics, and considering the cosmological constant and
nonlinear parameter as thermodynamic quantities which can vary, it
was shown that indeed there is a complete analogy  between the
nonlinear dilaton black holes with Van der Waals liquid-gas system
\cite{Dayyani3}. In all mentioned above, one assumes the charge of
the black hole as an external fixed parameter and treats the
cosmological constant as the pressure of the system which can
vary.

\begin{figure}\label{Fig1}
\centering \subfigure[ Equation of state for Van der Walls
liquid-gas system with $a=1$ and
$b=1$.]{\includegraphics[scale=0.4]{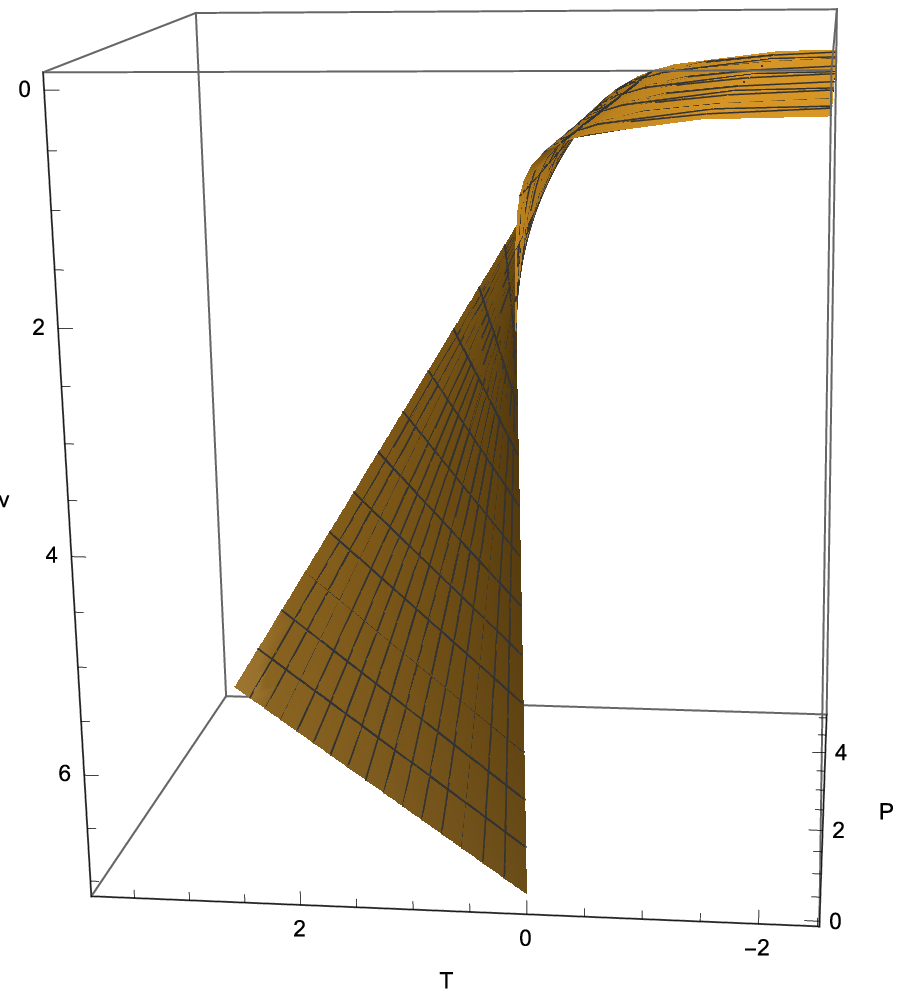}\label{Fig1a}}
\hspace*{.2cm} \subfigure[Equation of state of dilaton lifshitz
black hole with $Q=k=b=p=1$, $n=3$ and $z=1.1$.]
{\includegraphics[scale=0.4]{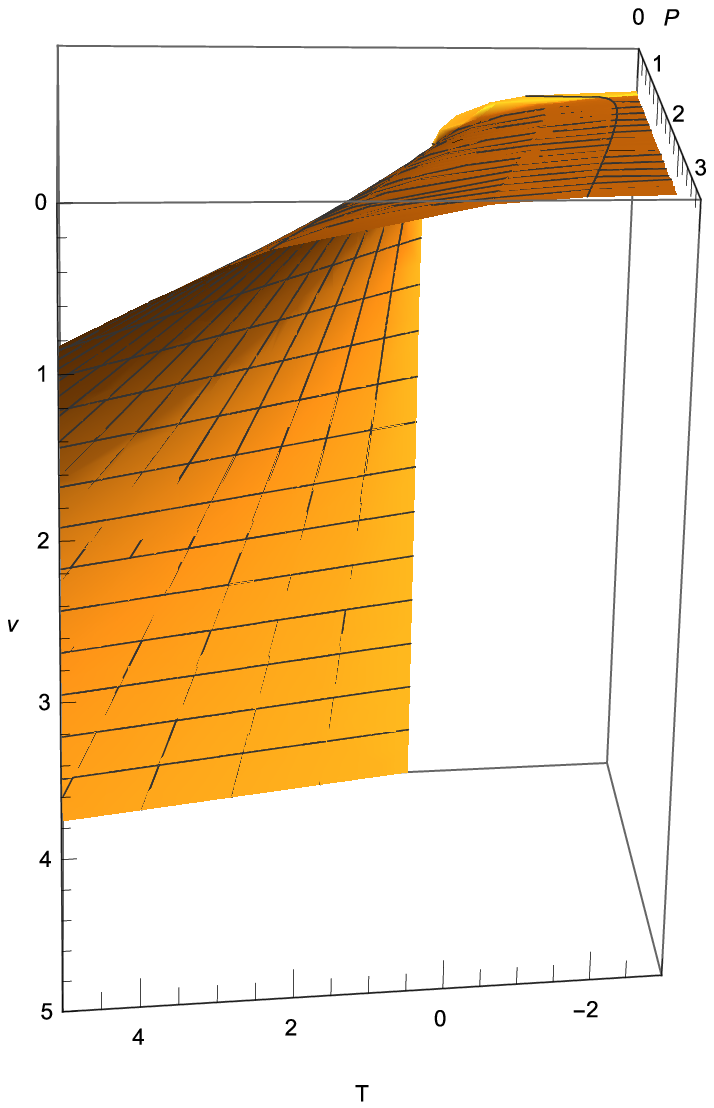}\label{Fig1b}}\caption{The 3-d
diagrams of equations of state for Van der Walls fluid system and
Lifshitz black holes. Comparing two diagrams indicate that these
systems have similar phase transition.}
\end{figure}

In the present work, we would like to investigate the critical
behaviour of Lifshitz black holes in Einstein-dilaton gravity in
the presence of a power-law Maxwell field. It is worthwhile to
mention that the $(n+1)$-dimensional Lagrangian in power Maxwell
theory is conformally invariant provided $p=(n+1)/4$ where $p$ the
power of the Lagrangian. Let us first plot a $3$-dimensional
diagrams for equation of state of Lifshitz black hole to
understand the phase behaviour of this system and show its analogy
with Van der Walls liquid-gas system (see Fig. \ref{Fig1}).
A close look at the temperature expression of Lifshitz-dilaton
black holes (see \cite{Kord5} and Eq. (\ref{Temp}) of the present
paper), shows that it is nearly impossible to solve this equation
for $P$ (or more precisely for $l$). Therefore, we cannot have an
analytical equation of state, $P=P(v,T)$, to investigate the
critical behavior or calculate critical quantities of Lifshitz
black holes. Another way to investigate critical behavior of the
black holes is to use the method of Refs. \cite{Cha1,Cha2}, but as
shown in \cite{Dehy}, such a view of thermodynamic conjugate
variables ($Q$ and $\Phi =Q/r_{+}$) which are not mathematically
independent can lead to physically irrelevant quantities such as
$(\partial Q/\partial \Phi )_{T}$ which is supposed to be a
thermodynamic response function, but mathematically ill-defined.

To address this problem, an alternative viewpoint toward
thermodynamic phase space of black holes was developed in
\cite{Dehy} by treating the cosmological constant as a fixed
parameter and considering the charge of the black hole as a
thermodynamic variable \cite{Amin1,Amin3}. It was argued that,
with fixed cosmological constant, the critical behavior indeed
occurs in $Q^{2}$-$\Psi$ plane, where $\Psi =1/2r_{+}$ is
conjugate of $Q^2$, and thus the equation of state is written as
$Q^{2}=Q^{2}(T,\Psi)$. We find out that in case of Lifshitz
dilaton black holes, the system admits a critical behaviour
provided we take the electrodynamics in the form of power-Maxwell
field and considering $Q^{s}$ as a thermodynamic variable with
$\Psi=\left({\partial M}/{\partial Q^{s} }\right)_{S,P}$ as its
conjugate, where $s=2p/(2p-1)$. In this case we can define a new
response function which naturally leads to physically relevant
quantity. Thus, the equation of state is written in the form of
$Q^{s}=Q^{s}(T,\Psi)$ and Smarr relation based on this new phase
space as $ M=M(S,Q^{s},P)$. Clearly, for $p=1$, the power Maxwell
field reduces to standard Maxwell field and $Q^{s}\rightarrow
Q^2$. Following \cite{Dehy}, in this approach we keep the
cosmological constant (pressure) as a fixed quantity, while the
charge of the system can vary.

This paper is outlined as follows. In the next section, we present
the action, basic field equations of of Lifshitz dilaton black
holes and review thermodynamic properties of this system. In
section \ref{Structure}, we study the phase structure of the
solution and present the modified Smarr relation. In section
\ref{eq of state}, we obtain the equation of state and study the
critical behavior of the solutions and compare them with Van der
Waals fluid system. We investigate the Gibbs free energy and the
critical exponents of the system in sections \ref{Gibbs} and
\ref{Exponent}, respectively. The last section is devoted to
summery and conclusion.
\section{Thermodynamics of Lifshitz dilaton black holes}\label{Field}
In this section we are going to review the solutions of charged
Lifshitz black holes with power Maxwell field \cite{Kord5}, with
emphasizing on their thermodynamic properties. The
$(n+1)$-dimensional action of Einstein-dilaton gravity in the
presence of a power Maxwell electromagnetic and two linear Maxwell
fields can be written as
\begin{eqnarray}\label{action}
S &=&-\frac{1}{16\pi }\int_{\mathcal{M}}d^{n+1}x\sqrt{-g}\left\{ \mathcal{R}-%
\frac{4}{n-1}(\nabla \Phi )^{2}\right.  \notag  \label{action2} \\
&&\left. -2\Lambda +\left( -e^{-4/(n-1)\lambda _{1}\Phi }F\right)
^{p}-\sum\limits_{i=2}^{3}e^{-4/(n-1)\lambda _{i}\Phi
}H_{i}\right\},
\end{eqnarray}
where $\mathcal{R}$ is the Ricci scalar on manifold $\mathcal{M}$,
$\Phi $ is the dilaton field, $\lambda _{1}$ and $\lambda _{i}$
are constants. In Eq. (\ref{action2}) $F=F_{\mu \nu} F^{\mu \nu}$
and $H_{i}$ are the Maxwell invariants of electromagnetic fields,
where $F_{\mu \nu }=\partial _{\lbrack \mu }A_{\nu ]}$ and $\left(
H_{i}\right) _{\mu \nu }=\partial _{\lbrack \mu }\left(
B_{i}\right) _{\nu ]}$, with $A_{\mu }$ and $\left( B_{i}\right)
_{\mu }$ are the electromagnetic potentials. Varying the action (\ref%
{action}) with respect to the metric $g_{\mu \nu }$, the dilaton field $\Phi
$, electromagnetic potentials $A_{\mu }$ and $\left( B_{i}\right) _{\mu }$%
, lead to the following field equations \cite{Kord5}
\begin{eqnarray}
&&\mathcal{R}_{\mu \nu }=\frac{g_{\mu \nu }}{n-1}\left\{ 2\Lambda
+(2p-1)\left( -Fe^{-4\lambda _{1}\Phi /(n-1)}\right) ^{p}\right.  \notag \\
&&\left. -\sum\limits_{i=2}^{3}H_{i}e^{-4\lambda _{i}\Phi /(n-1)}\right\} +%
\frac{4}{n-1}\partial _{\mu }\Phi \partial _{\nu }\Phi  \notag \\
&&+2pe^{-4\lambda _{1}p\Phi /(n-1)}(-F)^{p-1}F_{\mu \lambda }F_{\nu }^{\text{
\ }\lambda }  \notag \\
&&+2\sum\limits_{i=2}^{3}e^{-4\lambda _{i}\Phi /(n-1)}\left( H_{i}\right)
_{\mu \lambda }\left( H_{i}\right) _{\nu }^{\text{ \ }\lambda },
\label{FFE1} \\
&&\nabla ^{2}\Phi -\frac{p{\lambda _{1}}}{2}e^{-{4\lambda _{1}p\Phi }/({n-1}%
)}(-F)^{p}  \notag \\
&&+\sum\limits_{i=2}^{3}\frac{{\lambda _{i}}}{2}e^{-{4\lambda _{i}\Phi }/({%
n-1})}H=0, \\
&&\triangledown _{\mu }\left( e^{-{4\lambda _{1}p\Phi }/({n-1}%
)}(-F)^{p-1}F^{\mu \nu }\right) =0, \\
&&\triangledown _{\mu }\left( e^{-{4\lambda _{i}\Phi }/({n-1})}\left(
H_{i}\right) ^{\mu \nu }\right) =0.  \label{FFE4}
\end{eqnarray}%
We assume the line element of the higher-dimensional asymptotic
Lifshitz spacetime has the following form \cite{Kord5}
\begin{equation}
ds^{2}=-\frac{r^{2z}f(r)}{l^{2z}}dt^{2}+{\frac{l^{2}dr^{2}}{r^{2}f(r)}}%
+r^{2}d\Omega _{n-1}^{2},  \label{metric}
\end{equation}%
where $d\Omega _{n-1}^{2}$ is an ($n-1$)-dimensional hypersurface
with constant curvature $(n-1)(n-2)k$ and volume $\omega _{n-1}$.
Following the method of \cite{Kord5}, one can find the solutions
of the field equations (\ref{FFE1})-(\ref{FFE4}) as
\begin{eqnarray}
f(r) &=&1-{\frac{m}{{r}^{n-1+z}}}+{\frac{k{l}^{2}\left( n-2\right) ^{2}}{%
\left( z+n-3\right) ^{2}{r}^{2}}}+\frac{q^{2p}}{{r}^{\Gamma +z+n-1}},  \notag
\\
&&  \label{ff} \\
{\Phi (r)} &{=}&\frac{(n-1)\sqrt{z-1}}{2}{\ln \left( \frac{r}{b}\right) ,} \\
\left( A_{1}\right) _{t} &=&-\frac{q_{1}{b}^{2(z-1)}}{\Gamma {r}^{\Gamma }},\label{A1}
\\
\left( A_{2}\right) _{t} &=&\sqrt{\frac{z-1}{2(n+z-1)}}{\frac{{r}^{n+z-1}}{{l%
}^{z}{b}^{n-1}}}, \\
\left( A_{3}\right) _{t} &=&{\frac{\sqrt{{k\left( n-1\right) \left(
n-2\right) \left( z-1\right) }}{r}^{z+n-3}\,}{\sqrt{2}(z+n-3)^{3/2}{l}^{z-1}{%
b}^{n-2}}},
\end{eqnarray}%
where
\begin{eqnarray}
\Gamma &=&z-2+(n-1)/(2p-1), \\
q^{2p} &=&\frac{\left( 2\,p-1\right) {b}^{2(z-1)}}{\left( n-1\right) {l}%
^{-2\,p\left( z-1\right) -2}\Gamma }\left( 2q_{1}^{2}\right)
^{p}, \label{chparameter} \\
\Lambda &=&-\frac{(z+n-1)(z+n-2)}{2l^{2}}.
\end{eqnarray}%
It was argued in \cite{Kord5} that $p$ and $z$ are restricted as
\begin{equation}
\begin{tabular}{ll}
$\text{for }p<1/2,$ & $z-1>(n-2p)/(1-2p),$ \\
&  \\
$\text{for }1/2<p\leq n/2,$ & $\text{all }z(\geq 1)\text{ values are allowed,%
}$ \\
&  \\
$\text{for }p>n/2,$ & $z-1>(2p-n)/(2p-1).$%
\end{tabular}
\label{constraint2}
\end{equation}%
Using the modified BY formalism \cite{BY}, one can calculate the
mass of the solution per unit volume ${\omega _{n-1}}$ as \cite{Kord5}%
\begin{equation}
M=\frac{(n-1)m}{16\pi l^{z+1}},  \label{Mass}
\end{equation}%
where the mass parameter $m$ can be written in term of the horizon radius $%
r_{+}$ by using the fact that $f(r_{+})=0$. We find
\begin{equation}
{m(r_{+})}={r}_{+}^{z+n-1}+{\frac{k{l}^{2}\left( n-2\right) ^{2}{r}%
_{+}^{z+n-3}}{\left( z+n-3\right) ^{2}}}+\frac{q^{2p}}{{r}_{+}^{\Gamma }}.
\label{mr+}
\end{equation}%
One can also calculate the charge of the black hole by applying
the Gauss law
\begin{equation}
Q=\frac{\,{1}}{4\pi }\int r^{n-1}e^{-{4\lambda _{1}p\Phi /(n-1)}%
}(-F)^{p-1}F_{\mu \nu }n^{\mu }u^{\nu }d{\Omega },  \label{chdef}
\end{equation}%
where $n^{\mu }$ and $u^{\nu }$ are the unit spacelike and timelike normals
to the hypersurface of radius $r$ given as
\begin{equation*}
n^{\mu }=\frac{1}{\sqrt{-g_{tt}}}dt=\frac{l^{z}}{r^{z}\sqrt{f(r)}}dt,\text{
\ \ \ \ }u^{\nu }=\frac{1}{\sqrt{g_{rr}}}dr=\frac{r\sqrt{f(r)}}{l}dr.
\end{equation*}%
Using (\ref{chdef}), we obtain the charge per unit volume $\omega _{n-1}$ as
\begin{equation}
Q=\frac{2^{p-1}\left( q_{1}l^{z-1}\right) ^{2p-1}}{4\pi }.  \label{charge}
\end{equation}%
The electric potential $U$, measured at infinity with respect to horizon is
defined by
\begin{equation}
U=A_{\mu }\chi ^{\mu }\left\vert _{r\rightarrow \infty }-A_{\mu }\chi ^{\mu
}\right\vert _{r=r_{+}},  \label{Pot}
\end{equation}%
where $\chi =p\partial _{t}$ is the null generator of the horizon. Using (\ref{A1}), we can obtain
electric potential
\begin{equation}
U =\frac{pq_{1}b^{2(z-1)}}{\Gamma r_{+}^{\Gamma }}.  \label{elecpot}
\end{equation}%
The entropy of the black holes can be calculated by using the area
law of the entropy which is applied to almost all kinds of black
holes in Einstein gravity including dilaton black holes. Thus, the
entropy of our solutions per unit volume $\omega _{n-1}$ is
\begin{equation}
S=\frac{r_{+}^{n-1}}{4}.  \label{entropy}
\end{equation}%
The Hawking temperature can also be obtained as
\begin{eqnarray}
T_{+} &=&\frac{r_{+}^{z+1}f^{\prime }\left( r_{+}\right) }{4\pi l^{z+1}}
=\frac{1}{4\pi }\left\{ \frac{{(n-1+z){r}_{+}^{z}}}{l^{z+1}}+{\frac{%
k\left( n-2\right) ^{2}{r}_{+}^{z-2}}{l^{z-1}\left( z+n-3\right) }}-\frac{%
\Gamma q^{2p}}{l^{z+1}{r}_{+}^{\Gamma +n-1}}\right\} .  \notag \\
&&  \label{Temp}
\end{eqnarray}%
As one can see from expression (\ref{Temp}) it is nearly
impossible to solve this equation for $P$ (or more precisely for
$l$) and write an analytical equation of state, $P=P(v,T)$ for an
arbitrary Lifshitz exponent $z$. This implies that, for the
Lifshitz dialton black holes, one cannot investigate the critical
behavior of the system through an extended $P-v$ phase space by
treating the cosmological constant (pressure) as a thermodynamic
variable. However, as we shall see in the next section, it is
quite possible to investigate the critical behaviour of this
system through a new $Q^s-\Psi$ phase space and show its
similarity with Van der Waals fluid system.
\section{Critical behaviour of Lifshitz dilaton black holes}
 \subsection{\textbf{Phase structure}}\label{Structure}
It is now generally accepted that charged black holes in AdS
spaces allow critical behavior similar to the Van der Waals fluid
system, provided one treats the cosmological constant as a
thermodynamic variable (pressure) in an extended phase space
\cite{MannRN}. Also it has been shown in \cite{Dehy} that there is
deeper connection between charged AdS black holes and Van der
Waals fluid system. Indeed, it was argued that similar behavior
can be found without extending the phase space \cite{Dehy} by even
keeping the cosmological constant as a fixed parameter. The key
assumption in this picture,  is to treat the square of the charge
of black hole, $Q^{2}$, as a thermodynamic variable instead of
charge $Q$ \cite{Dehy}. Besides, the equation of state has been
written as $Q^{2}=Q^{2}(T,\Psi)$ where $\Psi=1/v$ (conjugate of
$Q^{2}$) is the inverse of the specific volume. With this new
picture, the authors completed analogy of charged AdS black holes
with Van der Waals fluid system with exactly the same critical
exponents. In this section, we would like to consider Lifshitz
dilaton black holes with power-law Maxwell field and investigate
the critical behavior as well as analogy with Van der Walls fluid
for this system.

The usual first law of thermodynamics in an extended phase space
is in the from of
\begin{equation}\label{firstlaw}
dM=TdS+VdP+UdQ.
\end{equation}
From this point of view the usual Smarr relation which obtained
from thermodynamic variables (\ref{elecpot})-(\ref{Temp}) and mass
(\ref{Mass}) can be written as
\begin{equation}\label{smar}
M=\frac{n-1}{z+n-3}TS+\frac{-2}{z+n-3}VP+\frac{2p-1}{2p}\left(1+\frac{\Gamma}{z+n-3}\right)
UQ.
\end{equation}
where
\begin{equation}
P=\frac{n\left( n-1\right) r_+^{z-1}}{16 \pi ^{z+1}},\  \  \quad V
= \int{4S dr_+}=\frac{r_+^n \omega_{n-1}}{n}.
\end{equation}
It was shown in \cite{Dehy} that by replacing term $U dQ$ in the
first law with term $\Psi dQ^{2}$, the system allows critical
behavior similar to the Van der Waals fluid system. First of all,
let us review the motivation of this selection. The well-known
thermodynamic quantities of AdS black holes are given
\cite{Dehy,MannRN}
\begin{equation}
M=\frac{r_+}{2}+\frac{Q^2}{2r_+}+\frac{r_+^3}{2l^2},
\end{equation}
\begin{equation}
T=\frac{1}{4\pi r_+}\left(1+\frac{3r_+^2}{l^2}-\frac{Q^2}{r_+^2}
\right), \label{TAdS}
\end{equation}
\begin{equation}
U=\frac{Q}{r_+}  \Longrightarrow   QU=\frac{Q^2}{r_+},
\end{equation}
and the usual Smarr formula is \cite{Dehy,MannRN}
\begin{equation}
M=2\left(TS-VP \right)+QU.
\end{equation}
It is clear that $M$, $T$ and the term $(QU)$ in Smarr formula are
proportional to the square of the charge of black hole $Q^2$.
Therefore, when pressure  ($\Lambda$) is a fixed parameter, $Q^2$
is the best choice as a new variable \cite{Dehy} and one may
replace $UdQ$ by $\Psi d^2Q$, where $\Psi={1}/{2 r_+}$ is the
conjugate of $Q^2$ \cite{Dehy}. Therefore, inspired by the
expression (\ref{TAdS}), one may write the equation of state as
\cite{Dehy}
\begin{equation}
Q^2(T, \Psi)=r_+^2+\frac{3 r_+^4}{l^2}-4 \pi r_+^3 T . \label{eq
stateDehy}
\end{equation}
As mentioned before, in case of constant $l$ (or $\Lambda$) the
equation of state (\ref{eq stateDehy}) leads to a critical
behavior similar to Van der Walls fluid system \cite{Dehy}. Now,
we are going to employ this approach for charged Lifshitz black
holes with power Maxwell field. Let us write thermodynamic
quantities in term of $Q$ (the charge of black hole) instead of
$q$ (charge parameter). It is a matter of calculation to show by
using Eqs. (\ref{chparameter}) and (\ref{charge}), that the mass
(\ref{Mass}), temperature (\ref{Temp}) and electric potential
(\ref{elecpot}) are the same as
 \begin{eqnarray}
M=\frac{(n-1)\omega _{n-1}}{16\pi l^{z+1}}
\left\{{r}_{+}^{z+n-1}+{\frac{k{l}^{2}\left( n-2\right)^{2}{r}
_{+}^{z+n-3}}{\left( z+n-3\right) ^{2}}}+\frac{(2p-1)b^{2z-2} l^2
\pi^s Q^s}{(n-1) \Gamma 2^{\frac{-5s}{2}}\omega_{n-1}^s
r_+^{\Gamma}} \right\}
\end{eqnarray}
\begin{eqnarray}
T_{+} =\frac{1}{4\pi }\left\{ \frac{{(n-1+z){r}_{+}^{z}}}{l^{z+1}}+{\frac{%
 k\left( n-2\right) ^{2}{r}_{+}^{z-2}}{l^{z-1}\left( z+n-3\right) }}-\frac{(2p-1)b^{2z-2} 2^{\frac{5s}{2}}
 \pi^s Q^s}{(n-1)l^{z-1}  \omega_{n-1}^s {r}_{+}^{\Gamma +n-1}}\right\} \label{Temp2}
\end{eqnarray}
\begin{equation}
U = \frac{pb^{2(z-1)}}{\Gamma l^{z-1} r_{+}^{\Gamma }}
\left(\frac{\pi Q}{2^{3-p}w}\right)^{\left(s-1\right)}
\Longrightarrow QU=\frac{pb^{2(z-1)}}{\Gamma l^{z-1} r_{+}^{\Gamma
}} \left(\frac{\pi }{2^{3-p}w}\right)^{\left(s-1 \right)} Q^s.
\label{elecpot2}
\end{equation}
where
\begin{equation}
 s=2p/(2p-1).
\end{equation}
It is clear that the charge of black hole appears as $Q^s$ in  the
above equations. This motivates us to choose the new thermodynamic
variable as $Q^s$ which can simplify all calculations. Besides,
from Eq. (\ref{Temp2}) we see that any other choice except $Q^s$,
make the equation of state so complicated which cannot be solved
analytically to investigate it`s critical behavior and critical
quantities. While selecting $Q^s$ as a new variable causes to a
simple and solvable equation of state. Fortunately, in the limit
of $p=z=1$ and $n=3$, $Q^s \longrightarrow Q^2$ i.e., our result
reduces to the one of \cite{Dehy}. As we will show in the next
sections, the system allows the critical behavior similar to the
Van der Walls fluid with fixed cosmological constant by replacing
$U dQ$ with $\Psi dQ^{s}$ . Thus, we write down the first law in
the form
\begin{equation}
dM=TdS+VdP+\Psi dQ^{s},
\end{equation}%
where  $ T=\left( {\partial M}/{\partial S
}\right)_{P,Q^{s}} $, $ V=\left( {\partial M}/{\partial P
}\right)_{S,Q^{s}} $ and the conjugate of $Q^{s}$ is
\begin{equation}\label{psi}
\Psi=\left(\frac{\partial M}{\partial Q^{s}
}\right)_{S,P}=\frac{\left( 2\,p-1 \right) {\omega
}^{\frac{1}{2p-1}}{b}^{(2\,z-2)}{\pi }^{\frac{1}{2p-1}}}{16\Gamma
l^{z-1} r^{\frac {\eta}{2\,p-1}}}.
\end{equation}
while $\eta=2\,pz+n-4\,p-z+1$. When $p=z=1$ and in $3$-dimensions,
the above definition for $\Psi$ reduces to $\Psi={1}/{(2r_+)}$
\cite{Dehy} . In this new picture we can write the Smarr formula
for the charged Lifshitz dilaton black hole as
\begin{equation}\label{newsmar}
M=\frac{n-1}{z+n-3}TS+\frac{-2}{z+n-3}VP+\frac{2p-1}{2p}\left(1+\frac{\Gamma}{z+n-3}\right)
\Psi Q^{s}.
\end{equation}
It is worth noting that we have replaced the usual $\Phi dQ$ term
in
the first law with $\Psi dQ^{s}$. The extended phase space associated with $%
P=-\Lambda /(8\pi )$ is still the same. Using (\ref{elecpot2}) and
(\ref{psi}), straightforward calculations shows $QU=\Psi Q^s$ and
so Smarr equation (\ref{smar}) and (\ref{newsmar}) are the same.
\subsection{\textbf{Equation of state}}\label{eq of state} Using Eq.
(\ref{psi}) and treating pressure or more precisely $l$ as a fixed
parameter, Eq. (\ref{Temp}) can be written as
\begin{eqnarray}\label{eq state}
Q^{s}={\frac { \left( z+n-1 \right)
{Y}^{2\,\beta}}{X{l}^{z+1}}{\Psi}^{ {\frac {-2 \left( 2\,p-1
\right) \beta}{\eta}}}}+{\frac {k \left( n-2
 \right) ^{2}{Y}^{2\,\delta}}{ \left( z+n-3 \right) X {l}^{z-1}}{\Psi}
^{{\frac {-2 \left( 2\,p-1 \right) \delta}{\eta}}}}-{\frac {4 \pi
\,{\it T}}{X}{Y}^{{\frac {\alpha}{2\,p-1}}}{\Psi}^{-{\frac
{\alpha}{ \eta}}}},
\end{eqnarray}
where
\begin{eqnarray}\label{a b d e}
\alpha=2\,np+2\,pz-6\,p-z+2 ,\text{ \ \ \ \ \ \ }
\beta={\frac {np+2\,pz-3\,p-z+1}{2\,p-1}}, \notag \\
\delta={\frac {np+2\,pz-5\,p-z+2}{2\,p-1}},\text{ \ \ \ \ \ \ }
\eta=2\,pz+n-4\,p-z+1, \label{abde}
\end{eqnarray}
and
\begin{eqnarray}\label{XY}
X={\frac { \left( 2\,p-1 \right) {b}^{2\,z-2}{\pi
}^{s}{2}^{\frac{5s}{2}}}{{l} ^{z-1}{w}^{s} \left( n-1 \right)
}},\text{ \ \ \ \ \ \ } Y= \left( {\frac { \left( 2\,p-1 \right)
{b}^{2(z-1)}{2}^{\frac{5s}{2}} \pi^{\frac{s}{2p}}}{16
{l}^{z-1}\Gamma w^{\frac{s}{2p}}}} \right) ^{{\frac
{2\,p-1}{\eta}}},
\end{eqnarray}
In order to compare the critical behavior of the system with Van
der Waals gas, one may plot isotherm diagrams $Q^{s}-\Psi $, which
are displayed in Figs. \ref{Fig2}-\ref{Fig4}. As we know second
order phase transition occurs in the point with following
conditions:
\begin{figure}[tbp]
\epsfxsize=8cm \centerline{\epsffile{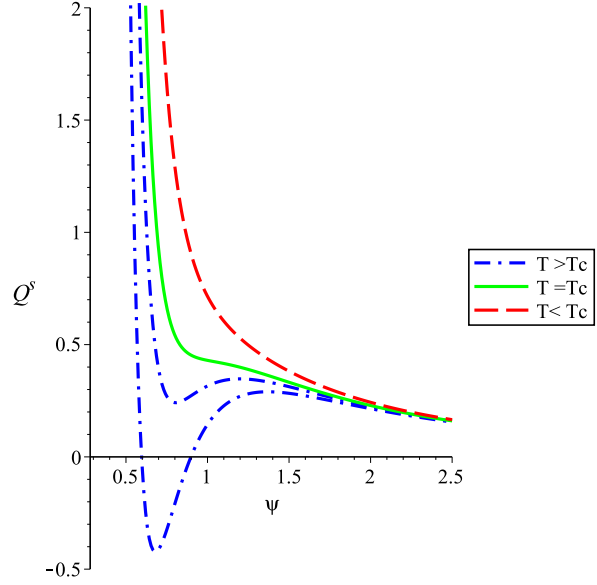}} \caption{$Q^s-
\Psi$ diagram of Lifshitz black holes for $b=1$, $n=3$, $q=1$,
$l=1$,
 $p =1$ and $z=0.6$.}
\label{Fig2}
\end{figure}

\begin{figure}[tbp]
\epsfxsize=8cm \centerline{\epsffile{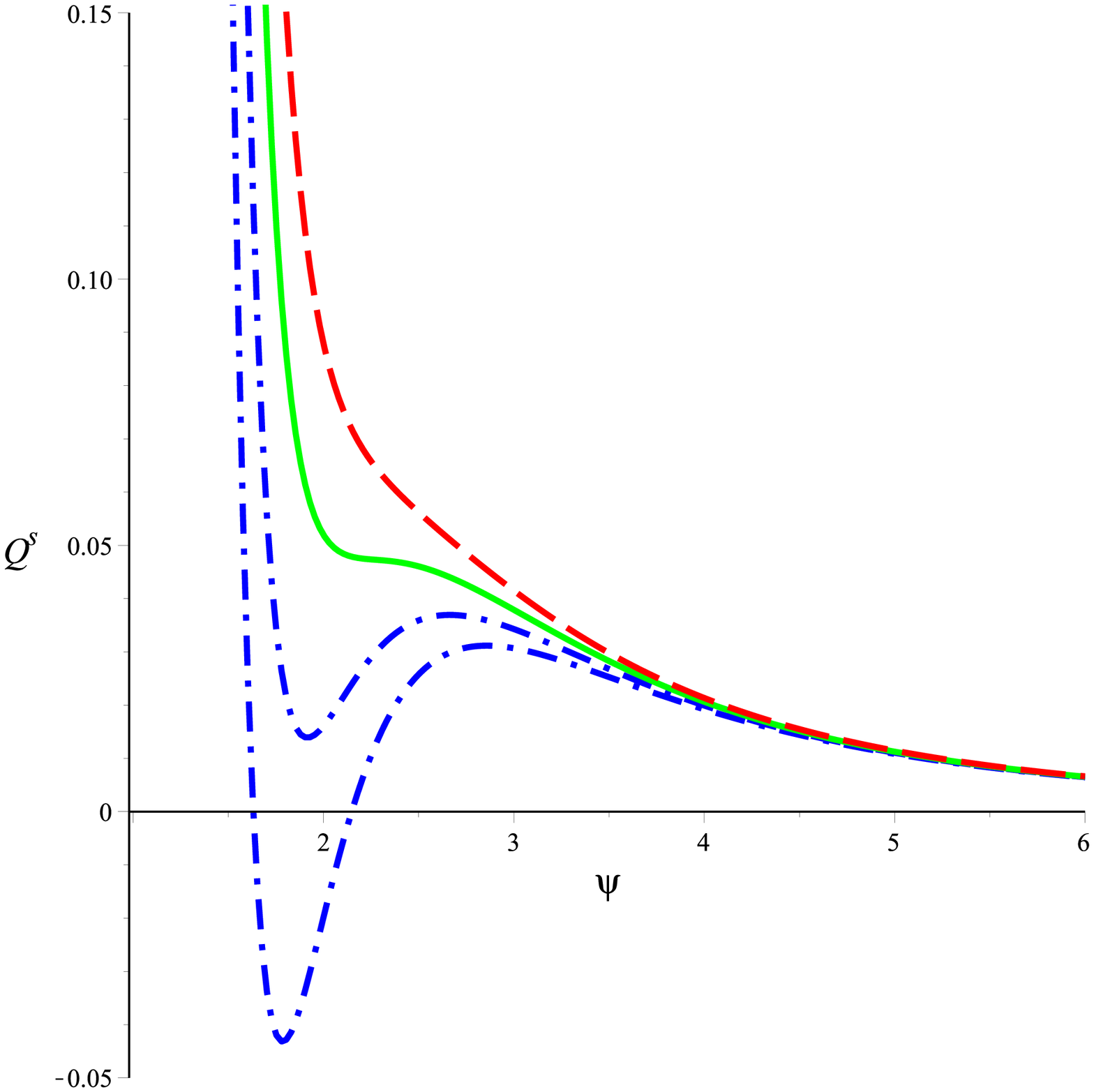}} \caption{$Q^s -
\Psi$ diagram of Lifshitz black holes for $b=1$, $n=3$, $q=1$,
$l=1$,
 $p =1.2$ and $z=1$.}
\label{Fig3}
\end{figure}

\begin{figure}[tbp]
\epsfxsize=8cm \centerline{\epsffile{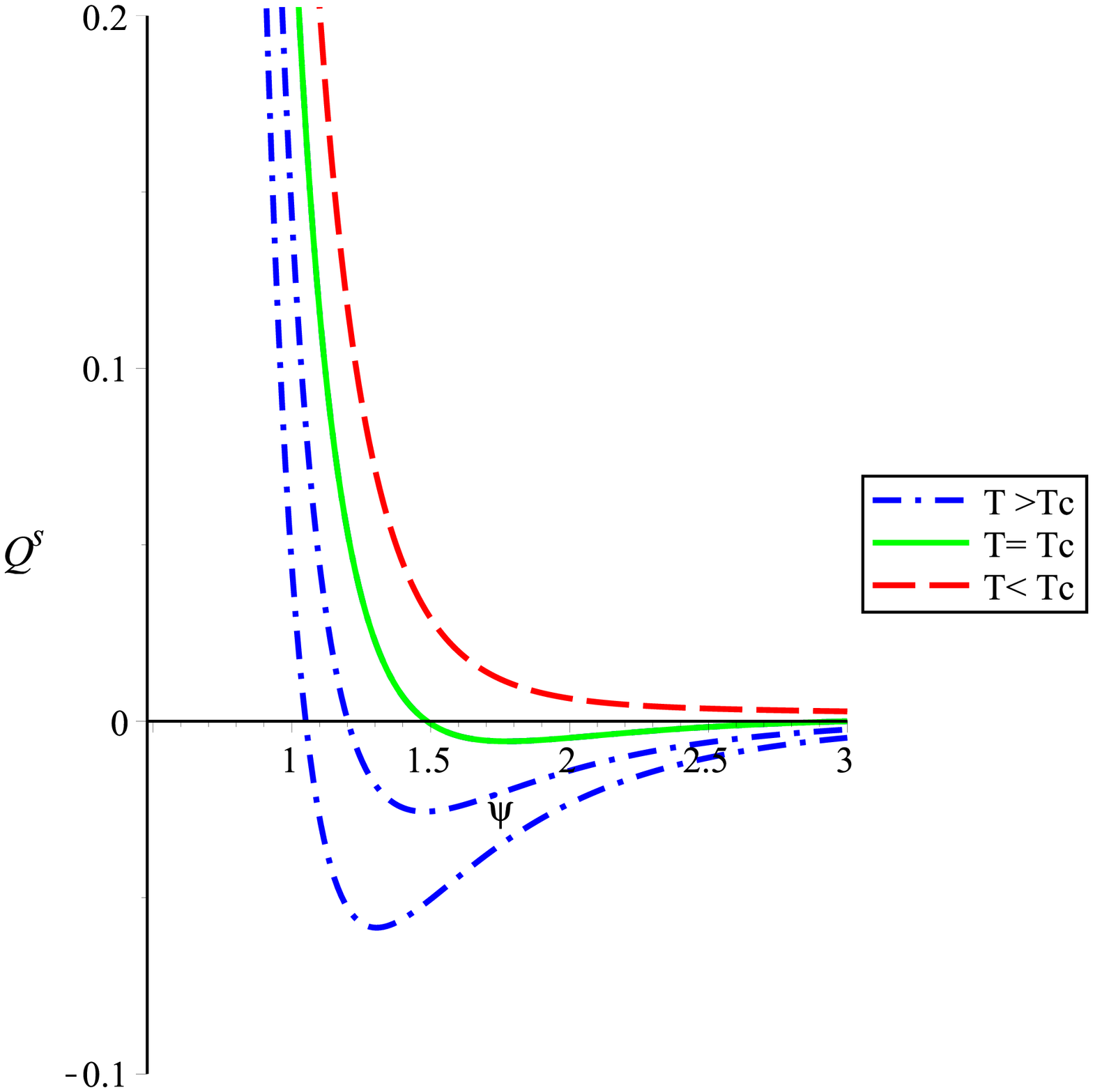}} \caption{$Q^s -
\Psi$ diagram of Lifshitz black holes for $b=1$, $n=3$, $q=1$,
$l=1$,
 $p =1.2$ and $z=1.4$.}
\label{Fig4}
\end{figure}
\begin{figure}
\centering \subfigure[$\rho_c -z$ for $b=1$, $n=3$, $q=1$
  and $p =1$.]{\includegraphics[scale=0.3]{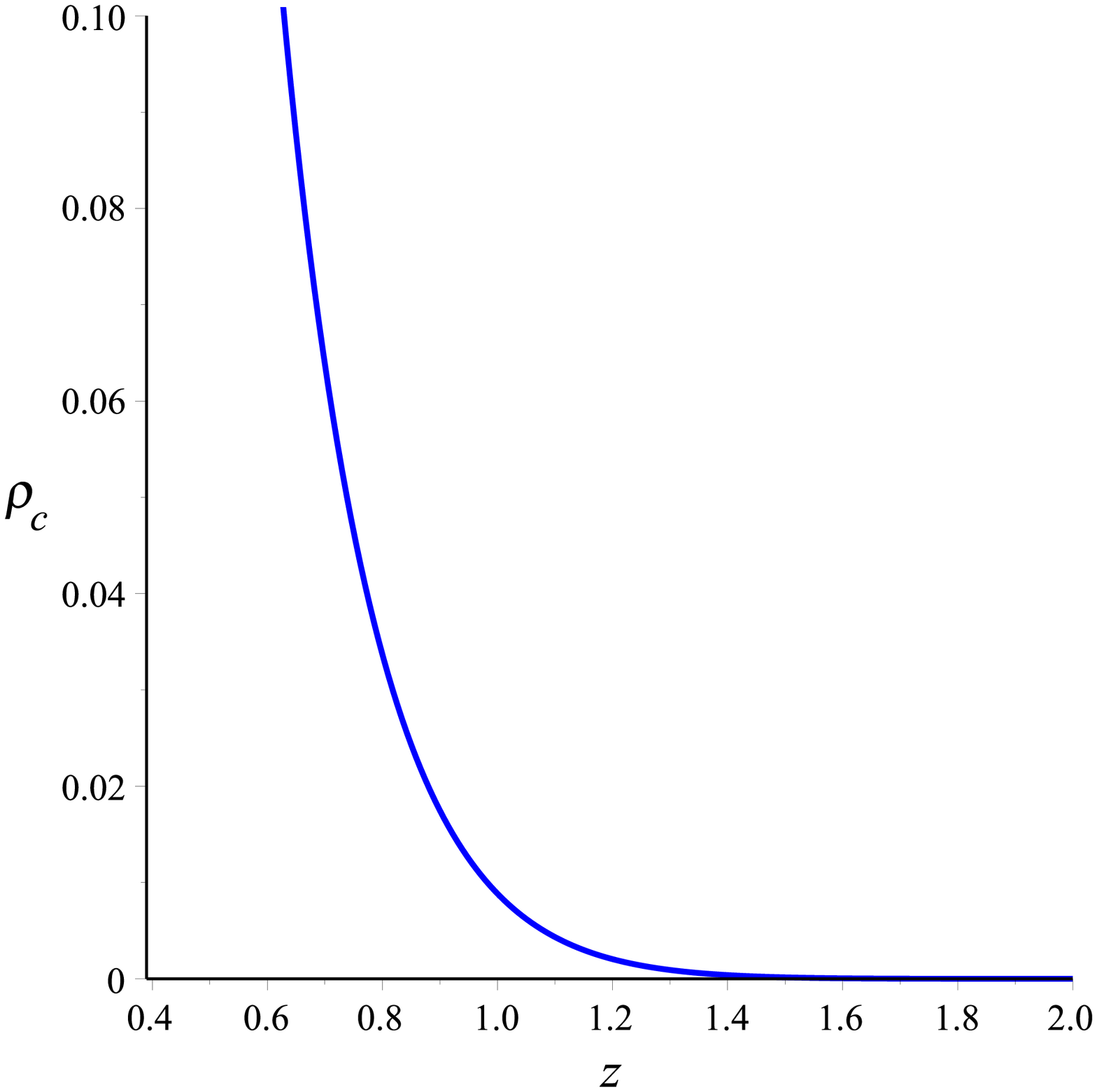}\label{Fig5a}}
\hspace*{.1cm} \subfigure[$\rho_c -p$ for $b=1$, $n=3$, $q=1$ and
$z =1$. As we see  $\rho_{c}$ is positive in the range $1/2< p
<3/2$.]{\includegraphics[scale=0.3]{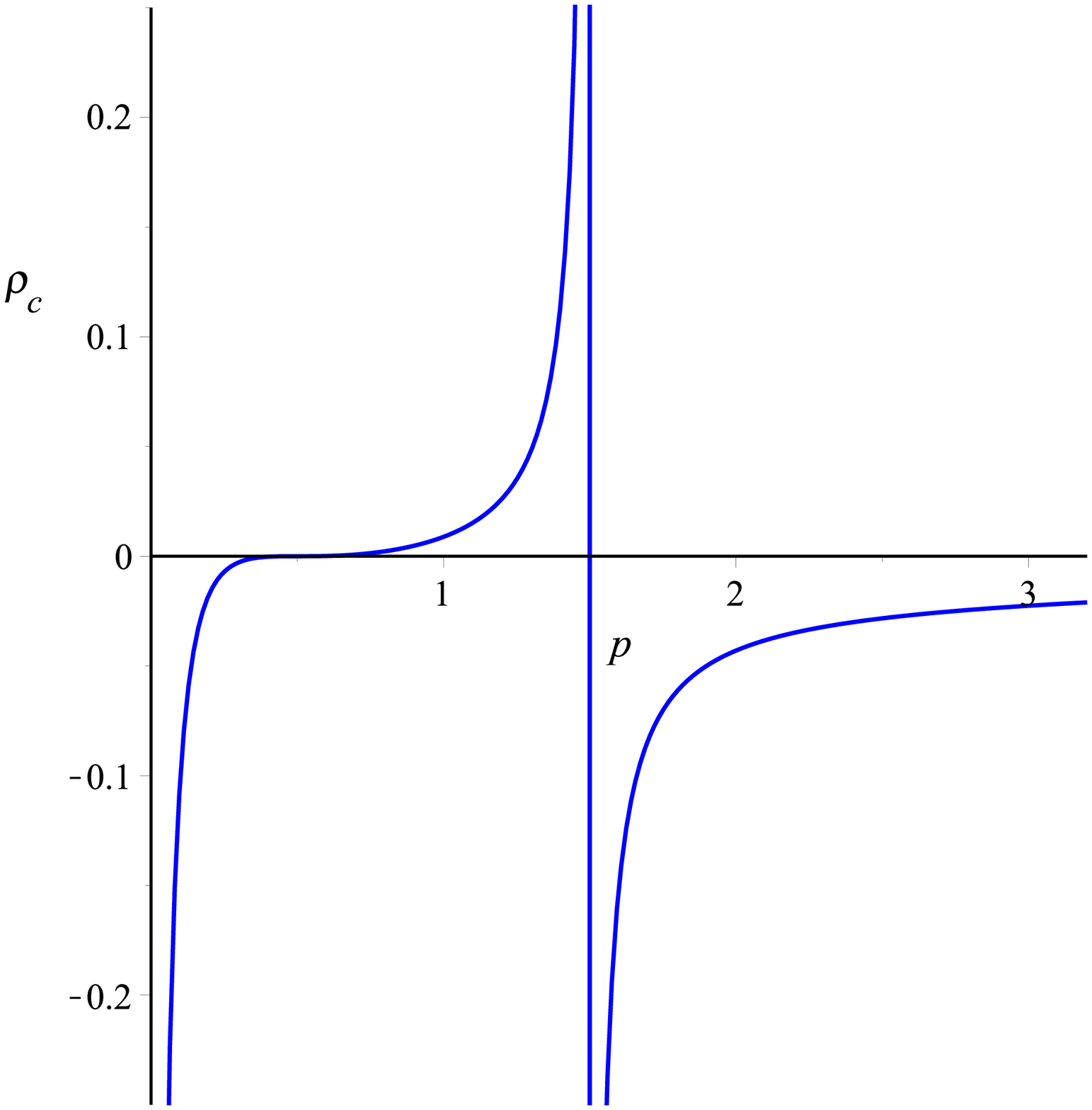}\label{Fig5b}}\caption{$\rho_C$
diagram of Lifshitz black holes.} \label{Fig5}
 \end{figure}
\begin{figure}
\centering \subfigure[$T_c$ versus $z$ for $b=1$, $n=3$, $q=1$ and
$p =1$]{\includegraphics[scale=0.3]{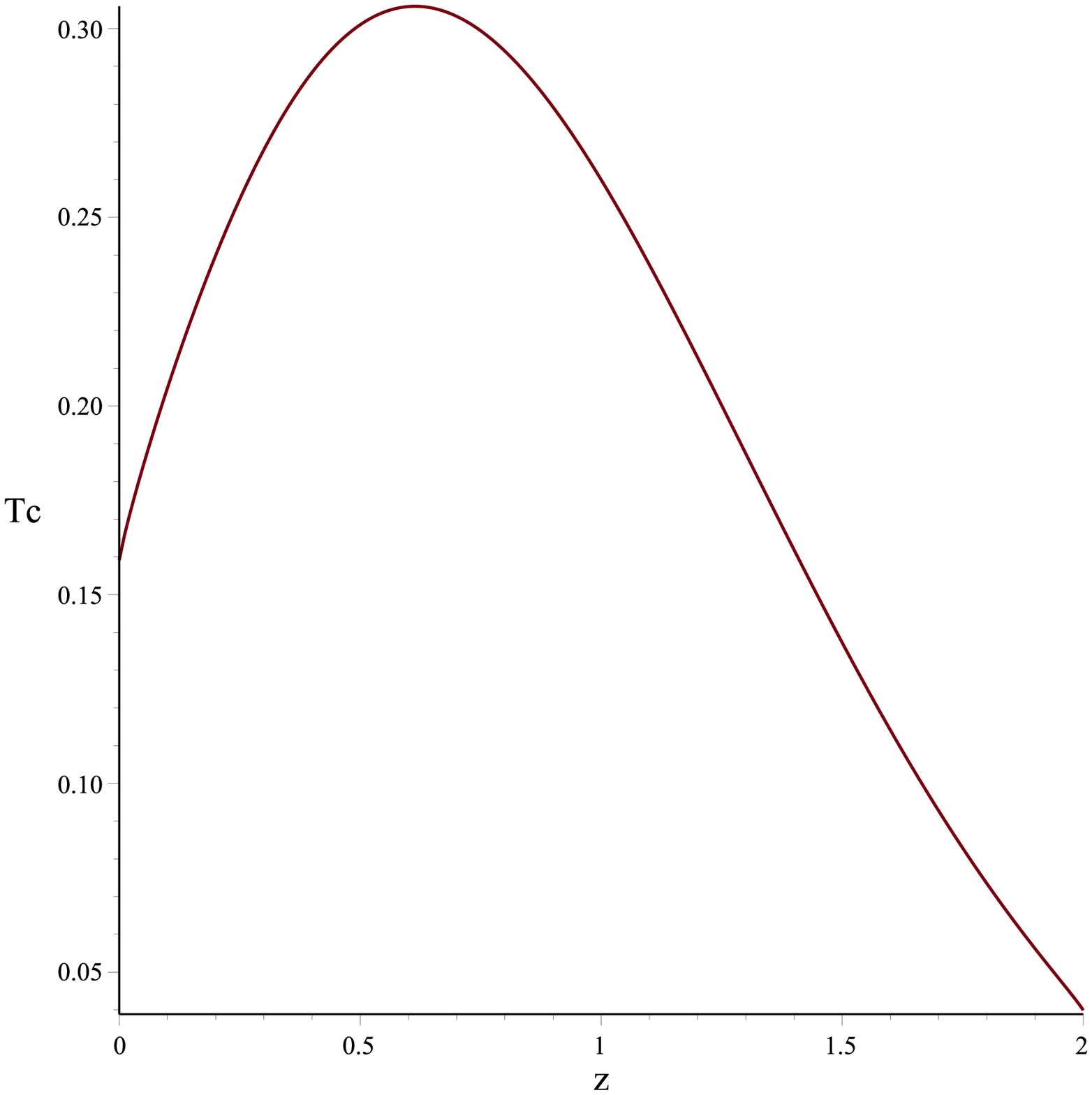}\label{Fig6a}}
\hspace*{.1cm} \subfigure[$T_c$ versus $p$ for $b=1$, $n=3$, $q=1$
 and $z =1$.]{\includegraphics[scale=0.3]{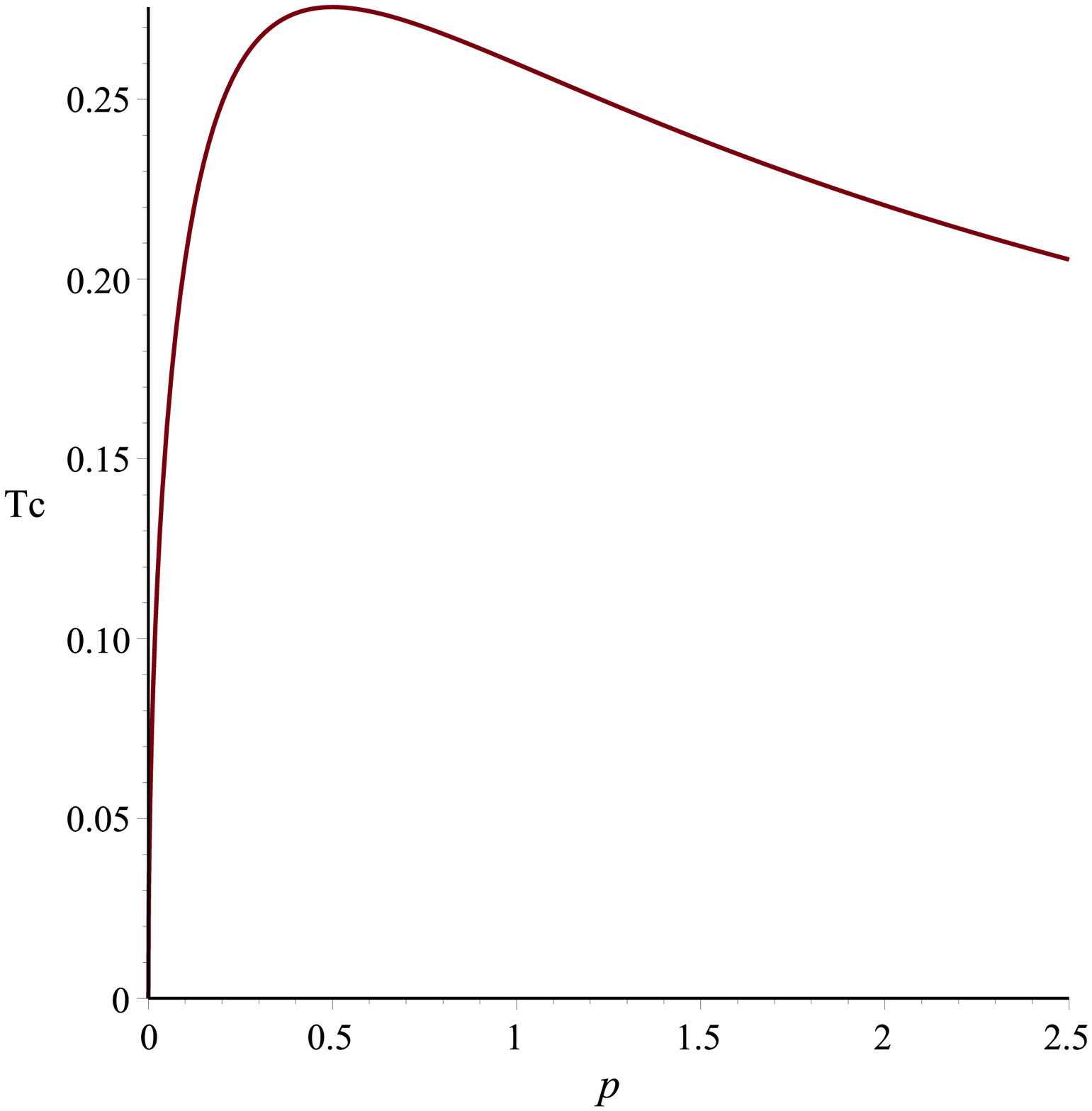}\label{Fig6b}}\caption{$T_{c}$ diagram of Lifshitz black holes.}
   \label{Fig6}
 \end{figure}
\begin{figure}
 \centering \subfigure[$\Psi_c$ versus $z$ for $b=1$, $n=3$, $q=1$
   and $p =1$]{\includegraphics[scale=0.3]{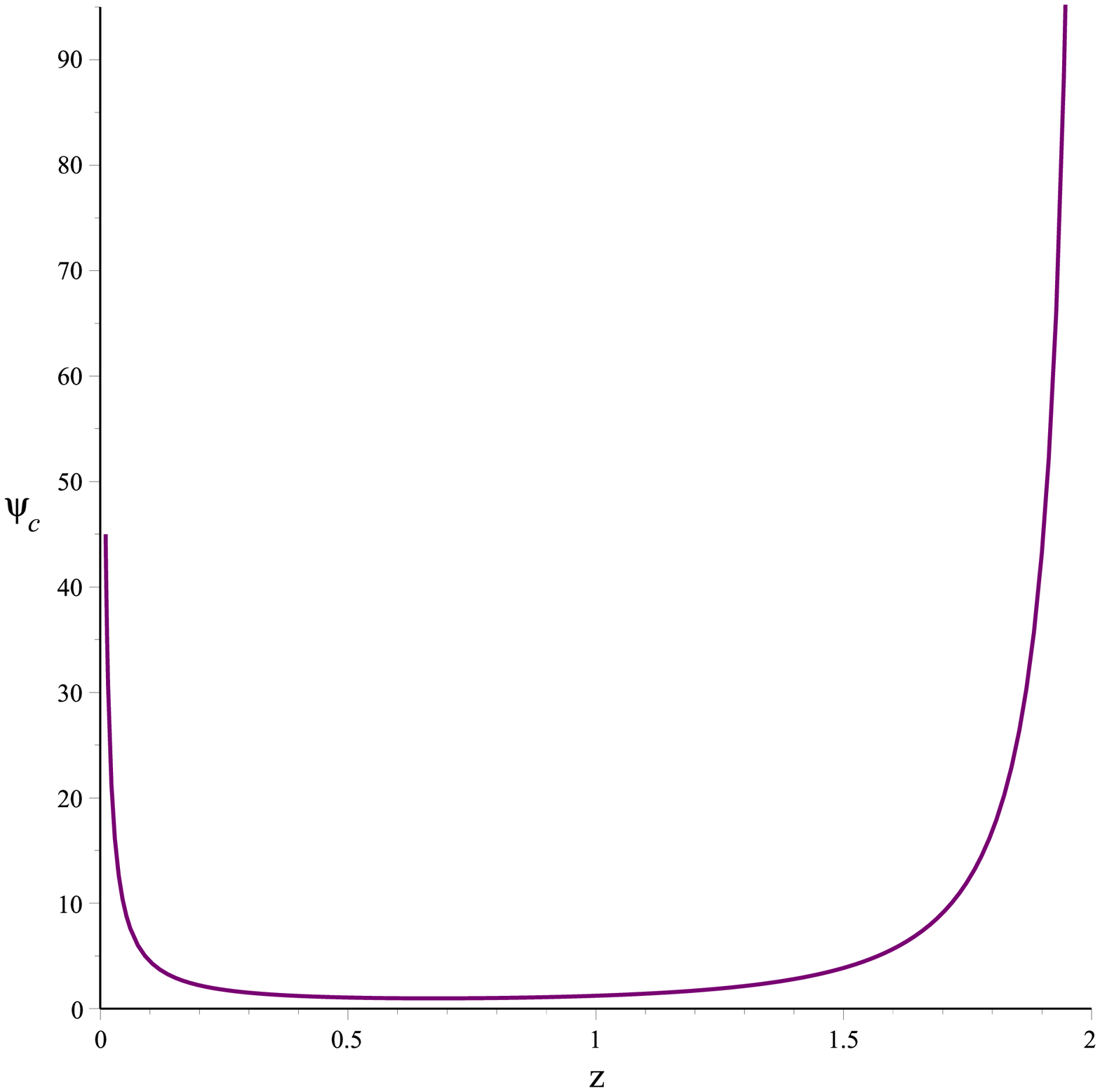}\label{Fig7a}}
  \hspace*{.1cm} \subfigure[$\Psi_c$ versus $p$ for $b=1$, $n=3$, $q=1$
    and $z =1$.]{\includegraphics[scale=0.3]{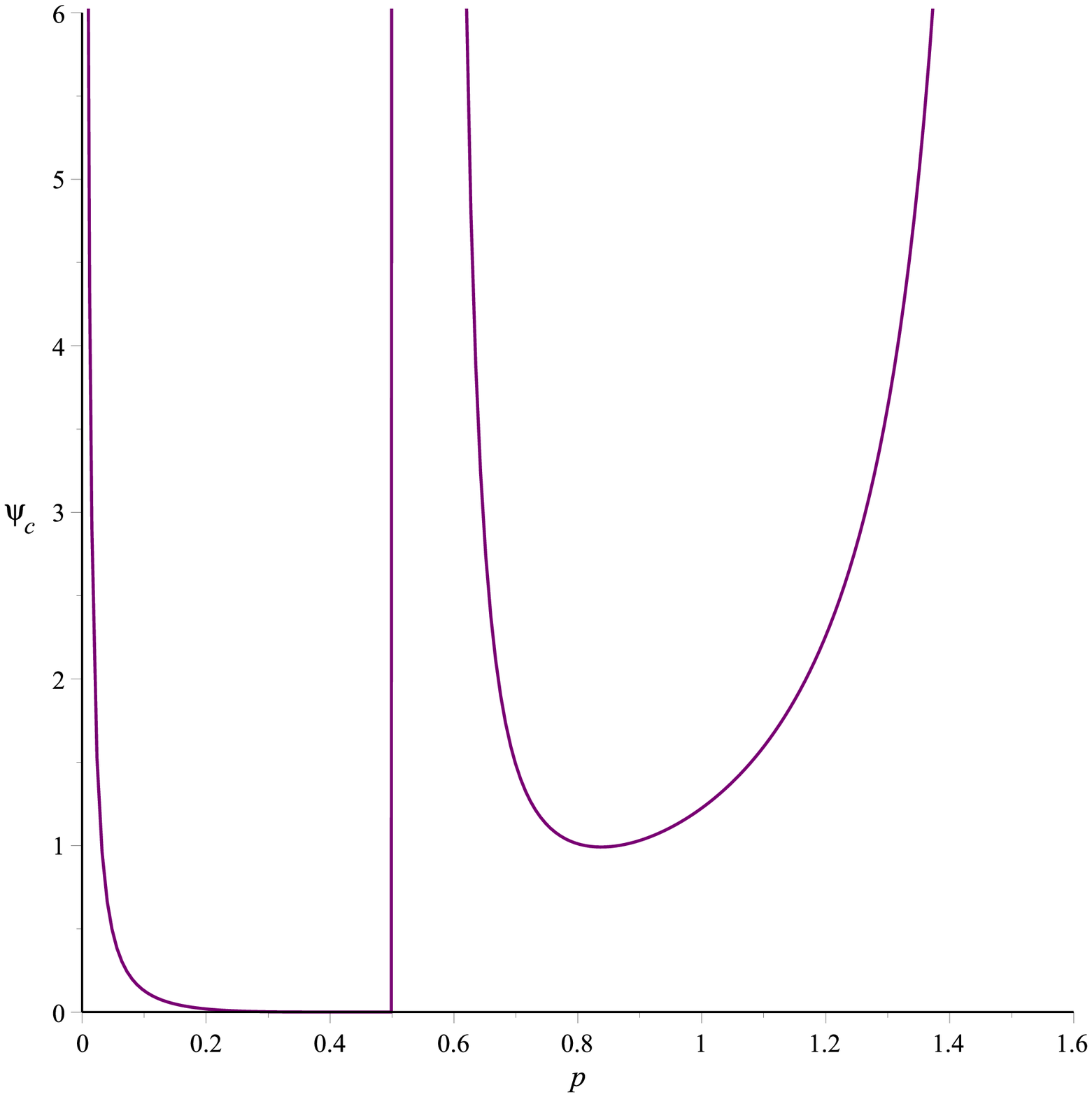}\label{Fig7b}}\caption{$\psi_{c}$ diagram of Lifshitz black holes.}
    \label{Fig7}
  \end{figure}
\begin{figure}
  \centering \subfigure[$Q^s_c$ versus $z$ for $b=1$, $n=3$, $q=1$
    and $p =1$]{\includegraphics[scale=0.3]{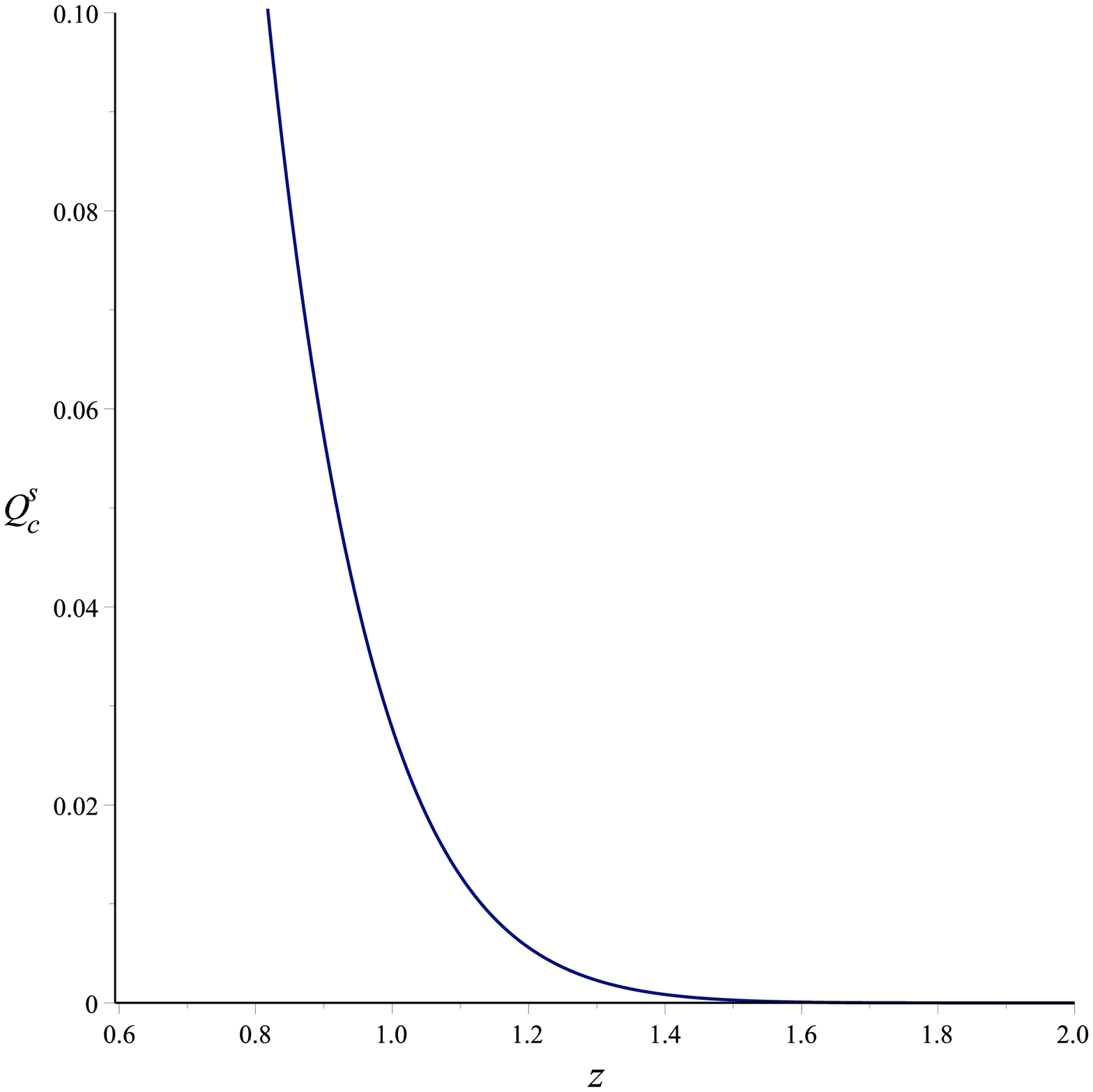}\label{Fig8a}}
   \hspace*{.1cm} \subfigure[$Q^s_c$ versus $p$ for $b=1$, $n=3$, $q=1$
     and $z =1$.]{\includegraphics[scale=0.3]{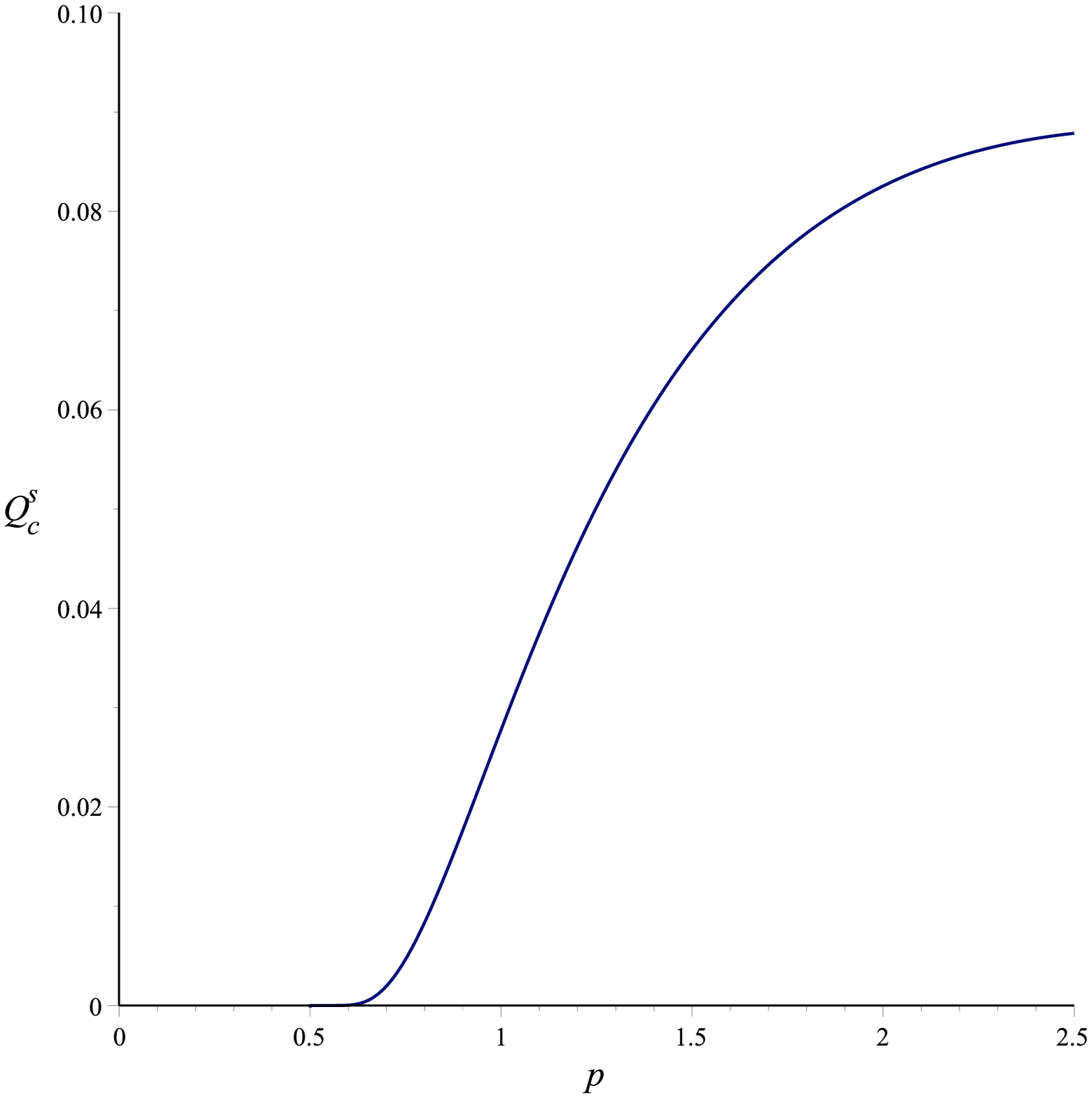}\label{Fig8b}}\caption{$Q^s_{c}$ diagram of Lifshitz black holes.}
     \label{Fig8}
   \end{figure}


\begin{equation}
\frac{\partial Q^{s}}{\partial \Psi}\Big|_{T_{c}}=0,\quad \frac{\partial ^{2} Q^{s}}{%
\partial \Psi^{2}}\Big|_{T_{c}}=0.  \label{CritEq}
\end{equation}
Solving Eqs. (\ref{CritEq}) yield the coordinates of the critical
point as
\begin{eqnarray}\label{Psi_c}
\Psi_{c}=\left( {\frac {k \left( n-2 \right) ^{2}\delta\,{l}^{2}
\left( -z+2 \right) }{{y}^{2} \left( z+n-3 \right) \beta\, \left(
z+n-1 \right) z }} \right) ^{{\frac {-\eta}{2(2\,p-1)}}},
\end{eqnarray}
\begin{eqnarray}\label{T_c}
T_{c}={\frac { \left( 2\,p-1 \right)  \left( n-2 \right) ^{z}}{l\pi \,\alpha
} \left( {\frac { \left( z+n-1 \right) \beta}{-z+2}} \right) ^{1-\frac{z}{2}}
 \left( {\frac {k\delta}{z \left( z+n-3 \right) }} \right)
 ^{\frac{z}{2}}},
\end{eqnarray}
\begin{eqnarray}\label{Q_c}
Q^{s}_{c}={\frac {w^{s} \left( -z+2 \right)  \left( 2\,p-1 \right) {b}^{-2\,z+2}
 \left( n-1 \right) }{2^{\frac{5s}{2}}\alpha \pi^{s}} \left( {\frac {{l}^{2} \left( -z+2
 \right)  \left( 2\,p-1 \right) \delta}{z \left( z+n-1 \right) }}
 \right) ^{\delta} \left( {\frac {k \left( n-2 \right) ^{2}}{ \left( z
+n-3 \right)  \left( 2\,p-1 \right) \beta}} \right) ^{\beta}},
\end{eqnarray}
Following the new definition $\rho_{c}=Q^{s}T_{c}\Psi_{c}$
\cite{Dehy}, the energy density of Lifshitz black hole at the
critical point is
\begin{eqnarray}\label{rho_c}
\rho_{c}={\frac { \left( 2p-1 \right) ^{2}w \left( n-1 \right)
 \left( n-2 \right) ^{n+2\,z-1} }{ {l}^{3-n}{
\pi }^{2}\Gamma\,{\alpha}^{2}}\left( z+n-1 \right) ^{\frac{5-n+2z}{2}} \left( {\frac { \left( -z+2 \right)
\delta}{z\beta}} \right) ^{\frac{n-3+2z}{2}} \left( {\frac {k}{z+n-3}}
 \right) ^{\frac{n-1+2z}{2}}}.
\end{eqnarray}
Since $\rho_{c}$ should be positive quantity, we can determine the
range of the parameters which satisfy this condition. In the
general case it is difficult to calculate it, but using diagram
(\ref{Fig5}), show that $\rho_{c}$ is positive
provided Eq. (\ref{constraint2}) is satisfied. For $n=3$,
$\rho_{c}$ is independent of the value of $l$.
As we expect when $n=3$, $z=1=p$, our results reduces to those of
RN-AdS black holes \cite{Dehy}
\begin{equation}
\Psi_c=\sqrt{\frac{3}{2l^2}}, \  \quad  Q^2_c=\frac{l^2}{36},
\quad T_c=\frac{1}{\pi l}\sqrt{\frac{2}{3}}, \quad \rho=
\frac{1}{36\pi}.
\end{equation}
To see how the critical quantities change with $p$ and $z$, one
may plot the Figs. \ref{Fig6}-\ref{Fig8}. It is clear that in the
limit of $z=2$, $\rho_{c}$  and $Q^{s}_{c}$ are equal to zero,
while $\Psi_{c}$ and $T_{c}$ goes to infinity. Indeed $T_{c}=0$ at
$z=2$, while for $z>2$, $T_{c}$ may have an imaginary value.

\subsection{\textbf{Gibbs free energy}}\label{Gibbs} 
\begin{figure}[tbp]
\epsfxsize=8cm \centerline{\epsffile{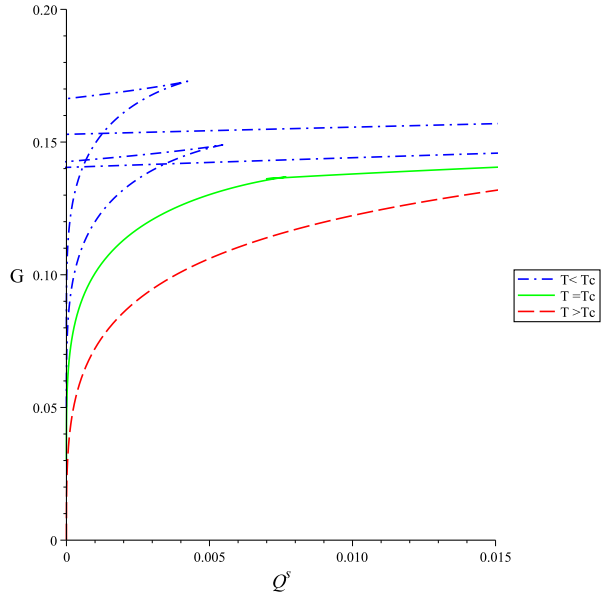}} \caption{$G- Q^s $
diagram of Lifshitz black holes for $b=1$, $n=3$, $q=1$, $l=1$,
 $p =.8$ and $z=1.4$.}
\label{Fig9}
\end{figure}

\begin{figure}[tbp]
\epsfxsize=8cm \centerline{\epsffile{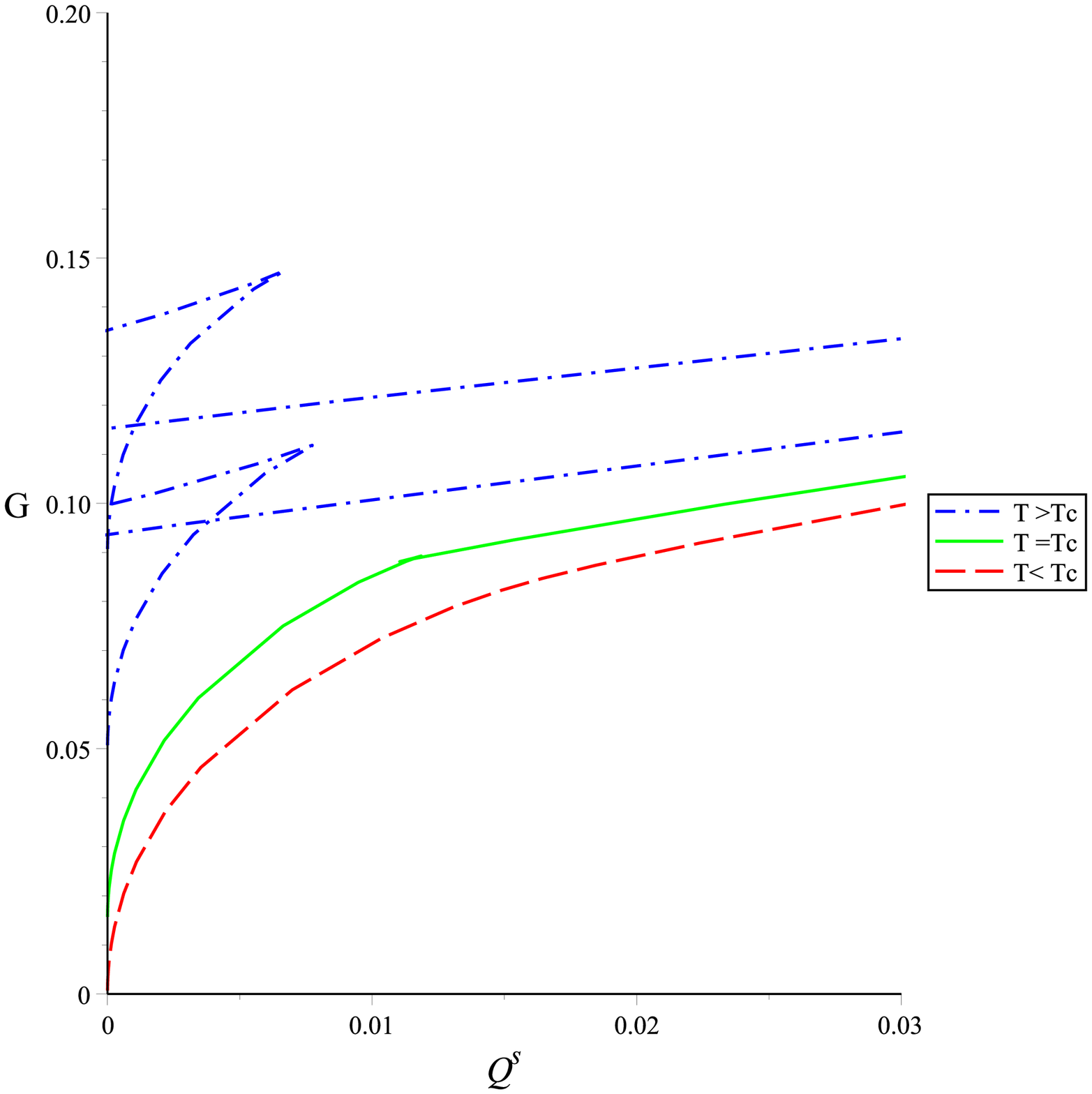}} \caption{$G- Q^s$
diagram of Lifshitz black holes for $b=1$, $n=3$, $q=1$, $l=1$,
 $p =1$ and $z=1.1$.}
\label{Fig10}
\end{figure}

\begin{figure}[tbp]
\epsfxsize=8cm \centerline{\epsffile{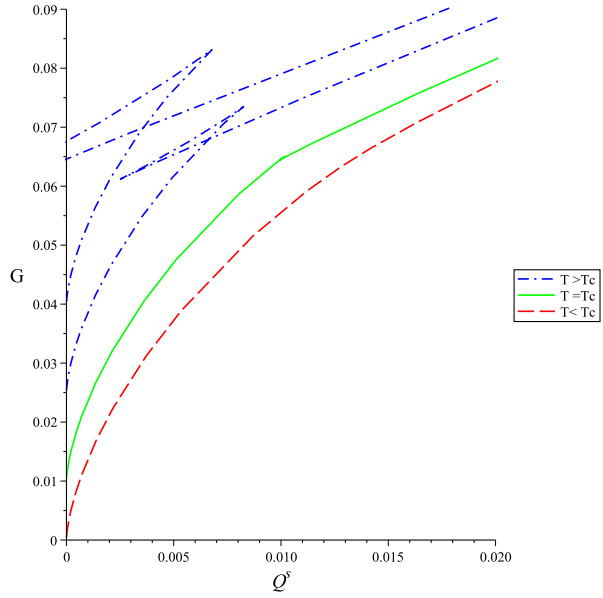}} \caption{$G- Q^s$
diagram of Lifshitz black holes for $b=1$, $n=3$, $q=1$, $l=1$,
 $p =1.2$ and $z=1.2$.}
\label{Fig11}
\end{figure}

Gibbs free energy is
one of the most important item which can help us to study phase
transition of a thermodynamical system. As we know, there is no
phase transition when Gibbs free energy is a continuous function.
Any discontinuity in Gibbs free energy, known as a zero order
phase transition. Also, first-order phase transition occurs when
the Gibbs free energy is continuous, but its first derivative with
respect to the temperature and pressure is discontinuous. At
first, we calculate the Gibbs free energy of Lifshitz dilaton
black hole. Then, we try to plot Gibbs diagrams to find out more
details about phase transition of the system. We associate the
energy of the system with the Gibbs free energy $G=M-TS$
\cite{Do1}. The Gibbs free energy can be obtained as
\begin{eqnarray}
G&=& G\left(Q^{s}, T\right)=  -\frac{k (n-2)^2 w (z-2) l^{1-z}
r^{n+z-3}}{16 \pi  (n+z-3)^2}-\frac{w l^{-z-1} (n+2 z-1)
r^{n+z-1}}{16 \pi } \nonumber \\
&&+\frac{2^{\frac{4-3 p}{2 p-1}} (2 p-1) \pi ^{\frac{1}{2 p-1}}
b^{2 z-2} l^{1-z} Q^{s} w^{\frac{1}{1-2 p}} (2 n p+n+2 p (2 z-5)-2
z+3) r^{\frac{n+2 p (z-2)-z+1}{1-2 p}}}{(n-1) \eta}.
\end{eqnarray}
In the limiting case where $p=z=1$ and $n=3$, the Gibbs free
energy reduces to \cite{Dehy}
\begin{equation}
G= G(T, Q^2)=
\frac{r_{+}}{4}+\frac{3Q^2}{4r_{+}}-\frac{r_{+}^3}{4l^2},
\end{equation}
where $r_{+}=r_{+}(T,Q^2)$. We have plotted the Gibbs energy
diagrams in Figs. \ref{Fig9}-\ref{Fig11}. These diagrams have been
shifted for more clarity. The swallowtail behavior of Figs.
\ref{Fig9}-\ref{Fig11} show that a first order phase transition
occurs in the system.

\subsection{\textbf{Critical exponents} \label{Exponent}} The behavior of
the physical quantities in the vicinity of critical point can be
characterized by the critical exponents. Following the approach of
\cite{MannBI}, one can calculate the critical exponents
$\alpha^{\prime }$, $\beta ^{\prime }$, $\gamma ^{\prime }$ and
$\delta ^{\prime }$ for the phase transition of charged Lifshitz
black holes in the presence of power-Maxwell field. To obtain the
critical exponents, we define the reduced thermodynamic variables
as
\begin{equation}
T_{r}=\frac{T}{T_c}\quad, \quad \psi_{r}=\frac{\psi}{\psi_c}
\quad, \quad Q^{s}_{r}=\frac{Q^{s}}{Q^{s}_{c}}.
\end{equation}
Since the critical exponents should be studied near the critical
point, we write the reduced variables in the form $T_r=1+t$ and
$\psi_r=1+\phi$, indicating deviation from the critical point. One
may expand Eq. (\ref{eq state}) near the critical point as
\begin{equation}
Q^{s}_{r}=1+At-Bt\phi -C\phi ^{3}+O\left( t\phi ^{2},\phi
^{4}\right), \label{ptw}
\end{equation}
where
\begin{equation}
A={\frac {-4\delta\,\beta}{z \left( -z+2 \right) }},\ \quad
B={\frac {-4\alpha\,\delta\,\beta}{z \left( -z+2 \right) \eta}},\
\  \  C={\frac {2 \left( 2\,p-1 \right)
^{2}\alpha\,\beta\,\delta}{3{\eta} ^{3}}}.
\end{equation}
To calculate the critical exponent $\alpha ^{\prime }$, we
consider the entropy $S$ given in Eq. (\ref{entropy}) as a
function of $T$ and $\psi$. Using Eq. (\ref{psi}) we have
\begin{equation*}
S=S\left( T,\Psi\right) ={\frac {{Y}^{n-1}}{4\pi }{\Psi}^{{
{(-2\,p+1)}/{\eta}}}}.
\end{equation*}%
Obviously, this is independent of $T$ and therefore the specific
heat vanishes, $C_{\psi}=T\left( \partial S/\partial T\right)
_{\psi}=0$. Since the exponent $\alpha ^{\prime }$ governs the
behavior of the specific heat at fixed $\psi$ $C_{v}\varpropto
\left\vert t \right\vert ^{\alpha ^{\prime }}$, hence the
exponent\ $\alpha ^{\prime }=0$.

Differentiating Eq.\ (\ref{ptw}) at a fixed $t<0$ with respect to
$\phi $, we get
\begin{equation}
dQ^{s}_{r}=-\left( Bt+3C\phi ^{2}\right) d\phi.
\end{equation}%
Now, we apply the Maxwell's equal area law\ \cite{callen}.
Denoting the variable $\phi$ for small and large black holes with
$\phi _{s}$ and $\phi _{l}$, respectively, we obtain
\begin{eqnarray}
Q^{s}_{r} &=&1+At-Bt\phi _{l}-C\phi _{l}^{3}=1+At-Bt\phi
_{s}-C\phi _{s}^{3},
\notag \\
0 &=&\int_{\phi _{l}}^{\phi _{s}}\phi dQ^{s}_{r}.  \label{Equal}
\end{eqnarray}%
Equation (\ref{Equal}) leads to the unique non-trivial solution
\begin{equation}
\phi _{l}=-\phi _{s}=\sqrt{-\frac{Bt}{C}},  \label{oml}
\end{equation}%
which gives the order parameter as
\begin{equation}
\left| \phi _{s}-\phi _{l}\right|  =2\phi
_{s}=2\sqrt{-\frac{B}{C}}\ t^{1/2}.
\end{equation}%
Thus, the exponent $\beta ^{\prime }$ which describes the
behaviour of the order parameter  near the critical point is
$\beta ^{\prime }=1/2.$ To calculate the exponent $\gamma ^{\prime
}$, one may determine the behavior of the following function near
the critical point
\begin{equation*}
\chi _{T}=\frac{\partial \psi}{\partial Q^s}\Big|_{T}.
\end{equation*}%
Differentiating Eq. (\ref{ptw}) with respect to $\phi$, near the
critical point may be written as
\begin{equation}
\chi _{T}\propto -\frac{%
\psi_{c}}{B Q^s_{c}}\frac{1}{t}\quad \Longrightarrow \quad \gamma ^{\prime }=1.
\end{equation}%
Finally, the shape of the critical isotherm $t=0$ is given by Eq.
(\ref{ptw}). We find
\begin{equation}
Q^{s}_{r}-1=-C\phi ^{3}\quad \Longrightarrow \quad \delta ^{\prime }=3.
\end{equation}%
\section{Summery and conclusions}
The critical behaviour of the Lifshitz dilaton black hole in an
extended phase space, where the cosmological constant is treated
as the thermodynamical pressure, cannot be studied due to the
complicated form of the solution. Indeed, it can be seen from Eq.
(\ref{Temp}) that it is almost impossible to solve this equation
for $P$ (or more precisely for $l$). Therefore, we cannot have an
analytical equation of state, $P=P(v,T)$, to investigate $P-v$
critically of the system. Also, investigating the phase space of
the system in $U-Q$ plan leads to physically irrelevant quantities
which are mathematically ill-defined \cite{Dehy}. Here, we
address, for the first time, the critical behavior of an
$(n+1)$-dimensional dilaton Lifshitz black hole in the presence of
a power-law Maxwell field via an alternative phase space developed
in \cite{Dehy}. We have treated the cosmological constant as a
fixed parameter and the charge of the system as thermodynamical
variable. It was argued in \cite{Dehy} that without extension the
phase space and by keeping the cosmological constant (pressure) as
a fixed quantity instead of the charge of the system, it is quite
possible to have critical behaviour similar to those of Van der
Walls system provided one take the equation of state of the form
$Q^{2}=Q^{2}(T,\psi)$ where $\Psi =1/2r_{+}$ is conjugate of
$Q^2$.

In this work, we disclosed that in order to investigate critical
behaviour of Lifshitz black holes with power Maxwell field, we
should modify the method developed in \cite{Dehy} by considering
$Q^{s}$ as a thermodynamic variable and write down the equation of
state in the form of $Q^{s}=Q^{s}(T,\psi)$, where $s=2p/(2p-1)$
with $p$ is the power of the power-Maxwell Lagrangian. In this
approach we keep the cosmological constant (pressure) as a fixed
quantity and treat the charge of the black hole as a thermodynamic
variable. This is in contrast to the extended phase space of
\cite{MannRN}, where the charge is fixed and the cosmological
constant is treated as a thermodynamic variable. The isotherm
diagrams $Q^{s}-\Psi$ show the complete analogy between our system
and Van der Walls liquid-gas system. Also the swallow tail
behavior of the Gibbs free energy represented a first order phase
transition is occurring in the system. Furthermore, we calculated
the critical quantities such as $T _c$, $\rho_c$, $P_c$ and
$\Psi_c$, at the critical point which depend on the metric
parameters. Finally, we obtained the critical exponents of the
system and found out that they are universal and exactly the same
as Van der Walls fluid system. Indeed, our study shows that the
approach here is powerful to investigate the critical behaviour of
Lifshitz black holes with power Maxwell field. It could help to
extract critical exponents of the system without extending the
phase space, which is useful in studying the thermodynamical
properties of the black holes. We expect to confirm that this
approach is viable and can be applied for other gravity theories
such as Gauss-Bonnet and Lovelock gravity.
\acknowledgments{We thank Shiraz University Research Council. The
work of AS has been supported financially by Research Institute
for Astronomy and Astrophysics of Maragha, Iran.}

\end{document}